\documentclass[prb,twocolumn,showpacs,preprintnumbers,amsmath,amssymb,showkeys]{revtex4}

\usepackage{graphicx}
\usepackage{dcolumn}
\usepackage{bm}

\begin{document}

\draft

\preprint{APS/123-QED}

\title{Coherent Diffraction Imaging of Single 95nm Nanowires}

\author{Vincent Favre-Nicolin}
 \altaffiliation[Also at ]{Universit\'e Joseph Fourier, Grenoble France}
 \email{Vincent.Favre-Nicolin@cea.fr}
\author{Jo\"el Eymery}%
\author{Robert K\"oster}%
\author{Pascal Gentile}%
\affiliation{CEA, INAC, F-38054 Grenoble, France}
\homepage{http://inac.cea.fr/sp2m/}

\date{\today}

\begin{abstract}
Photonic or electronic confinement effects in nanostructures become significant when one of their dimension is in the 5-300 nm range. Improving their development requires the ability to study their structure - shape, strain field, interdiffusion maps - using novel techniques. We have used coherent diffraction imaging to record the 3-dimensionnal scattered intensity of single silicon nanowires with a lateral size smaller than 100 nm. We show that this intensity can be used to recover the hexagonal shape of the nanowire with a $28\,nm$ resolution. The article also discusses limits of the method in terms of radiation damage.
\end{abstract}


\maketitle

  Vertical semiconductor nanowires (NWs) are developed as new, nanoscale building blocks for future electronic and photonic devices. For all applications, the properties depend strongly on the individual crystal structure and the average characteristics of the assembly of NWs. For some applications (resonant sensors, optical microcavities), device operation requires precise and matching characteristics (like diameters, longitudinal insertions or core/shell thicknesses) on many NWs. To characterize the structural properties of these new materials, transmission electron microscopy (TEM) is a standard tool that gives accurate information about individual objects when they are small (diameter $<20\,nm$).

\begin{figure}
\includegraphics[width=52mm,height=35mm]{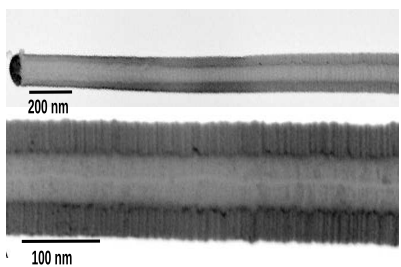}
\caption{\label{fig:fig1} Scanning electron microscopy images of a silicon nanowire, with an hexagonal-shaped section. The gold droplet used as a catalyst for the synthesis can be seen at the top of the wire (upper view). The bottom view shows the facets and nano-facetting for the same wire.}
\end{figure}

  X-ray diffraction is also a choice method to study nanostructures,\cite{stangl04}  particularly when chemical sensitivity (using anomalous scattering \cite{letoub04}) or strain mapping is required. But there is a limited number of X-ray diffraction studies on nanowires\cite{eymery07, mariager07,cham08}, due to the fact that for many samples it is not possible to perform a quantitative study on an \textit{assembly} of NWs: lack of an epitaxial relationship between the substrate and the vertical NWs, too large angular distribution of the wires, etc.. In this case it is required to conduct diffraction experiments on \textit{single NWs}. To this end, Coherent Diffraction Imaging (CDI) has been developed during the last few years :\cite{pfeifer06} in this method, scattering from a single object is recorded around one Bragg peak, and the recorded 3D data can be “inverted” to recover the shape of the scattering object. “State of the art” recent results show that it is now possible to study objects down to a few hundreds of nm in size, and more importantly that CDI allows to probe deformation inside a nanocrystal.\cite{pfeifer06} Existing studies have been focalised mostly on model materials with heavy atoms to yield a strong scattering of the X-ray beam (e.g. Pb and Au droplets).

   In the case of NWs, optical confinement effects begin to occur around 100-300 nm diameter (depending on the material), and electronic confinement around 5-30 nm.\cite{li06,pauz06} It is therefore of particular importance to study \textit{small} wires, with a diameter of the order of 100 nm. To study the possibility of using CDI for the study of sub-100 nm wires we conducted an experiment with homogeneous nanowires on the ID01 beamline of the European Synchrotron Research facility.

   The silicon NWs (see Fig.~\ref{fig:fig1}) were grown by low pressure (20 mBar) chemical vapor deposition at $650^{\circ}C$ via the gold-assisted Vapor-Liquid-Solid mechanism on (111) silicon substrate. 
 Au droplets defining the NW size were obtained by dewetting of 2D layers and silane (15 sccm) diluted in hydrogen was used as the reactive gas. The advantage of these Si NWs is that they are known to grow with little or no defects -particularly stacking faults along the $[111]$ direction- which is vital in a CDI experiment since it is required that all atoms diffract in a coherent manner. In the case of faulted NWs the scattered signal would be vastly different. \cite{cham08} $^,$\footnote{We have also studied the case of GaAs/GaP $[111]$ NWs which are known for their stacking faults along the growth direction. In such a case, the CDI signal recorded is essentially be a function of the faults nature and distribution along the NW axis, instead of the shape and elastic deformation of the wire.}

   During the experiment, the epitaxial wires were removed from the substrate and deposited on another silicon substrate, thus producing a random orientation for all the wires - this method decreases the likelihood that two different wires will diffract simultaneously on the detector for a given orientation of the sample, so that the diffraction of a single wire can be recorded. The sample was placed under He atmosphere to prevent oxidizing.

  The experiment was carried out using an X-ray photon energy equal to 10 keV, using beryllium compound refractive lenses \cite{snigirev98}  to focus the beam to $8.2\times13.6\,\mu m^2$ (larger than the NW lengths), with a photon flux equal to $4\times10^{10}\,ph/s$. The scattered intensity was measured using a Roper Scientific (SX-1300B) direct-illumination CCD, to obtain a maximum resolution and photon-counting efficiency.\cite{livet07} This detector presents $20\,\mu m$ pixels and was placed at 743 mm from the sample.

\begin{figure}
\includegraphics[width=52mm]{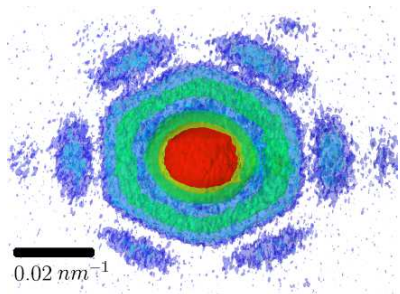}
\caption{\label{fig:fig2}(color online) 3D coherent diffraction image of a single Si $<111>$ nanowire, shown as a projection of multiple semi-transparent iso-surface layers, with the intensity increasing logarithmically from blue to green and red. The hexagonal symmetry of the wire is visible in the 3D diffraction image. The scale is given for $k=2\sin\theta/\lambda$.}
\end{figure}

  In order to collect the 3D scattered intensity around one Si $(111)$ Bragg peak, the detector was placed at the Bragg angle corresponding to the reflection, and the sample was then translated and/or rotated until one NW diffracted on the detector. Goniometer translations were used to ensure that the NW remains at the intersection of the X-ray beam and of the horizontal rotation axis. The complete scattering was then recorded by rotating the sample over a $1^{\circ}$ angular range with $0.01^{\circ}$ steps to yield a 3D pattern, which is then projected to an orthonormal frame of reference in reciprocal space (see Ref. \cite{vfn00} for details on the conversion between detector and reciprocal space coordinates) ; a view of the corrected 3D pattern is shown in Fig.~\ref{fig:fig2}.

  In the case of a single, non-strained nano-object subjected to an incoming coherent plane wave, the scattered amplitude $A$ is equal to :
\begin{eqnarray}
A({\mathbf{k}})=F(\mathbf{k}) FT[\Omega(\mathbf{r})]
\label{eq:eq1}
\end{eqnarray}

where $\mathbf{k}=\mathbf{k_f}-\mathbf{k_i}$ is the scattering vector near the $\mathbf{k_{Bragg}}$ Bragg position, $F(\mathbf{k})$ is the structure factor of a single unit cell, $FT[\Omega(\mathbf{r})]$ is the Fourier transform (FT) of the shape function $\Omega(\mathbf{r})$ of the nano-crystal, \textit{i.e.} a function which is equal to 1 inside the crystal and 0 outside.\footnote{Note that this approach is only correct as long as the crystal consists of complete unit cells - a reasonable approximation as long as the dimensions of the wire are much larger than the unit cell parameters.} As the structure factor is slowly varying near an existing Bragg peak, we can further assume that $F(\mathbf{k})\approx F(\mathbf{k_{Bragg}})$ and therefore that the scattered amplitude is the 3D FT of the nanocrystal's shape - or equivalently to its electronic density profile.

  During a CDI experiment only the square modulus of the amplitude is collected, and all phase information is lost. It is however possible to recover the lost phases using \textit{a priori} information on the object - in our case the inverse FT of the scattered amplitude corresponds to the projection of the electronic density of the sample and must be real, \textit{positive}, \textit{compact} and \textit{finite-sized}. These criterion have been used in a number of iterative phase retrieval algorithms \cite{gerchberg72,fienup82,marchesini03prb,wu05} to recover the phase corresponding to each point of the 3D reciprocal space.

\begin{figure}
\includegraphics[width=80mm]{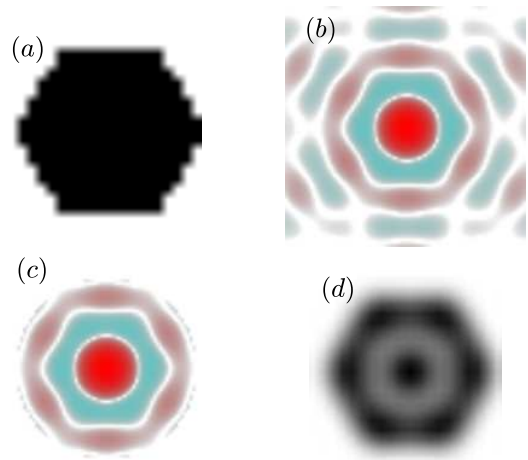}
\caption{\label{fig:figSIMU}(color online) Simulation of the truncation effect on the reconstruction of a NW shape. (a) original hexagonal shape of the NW ; (b) FT of the NW shape: as the original shape is centrosymmetric, the FT is real, here depicted with positive regions in red and negative in blue ; (c) truncation of the FT and (d) resulting NW shape calculated by inverse FT, in which dips (representing $\approx 15\%$ of the average density) can be clearly seen due to the truncation.}
\end{figure}

\begin{figure*}
\includegraphics[width=170mm]{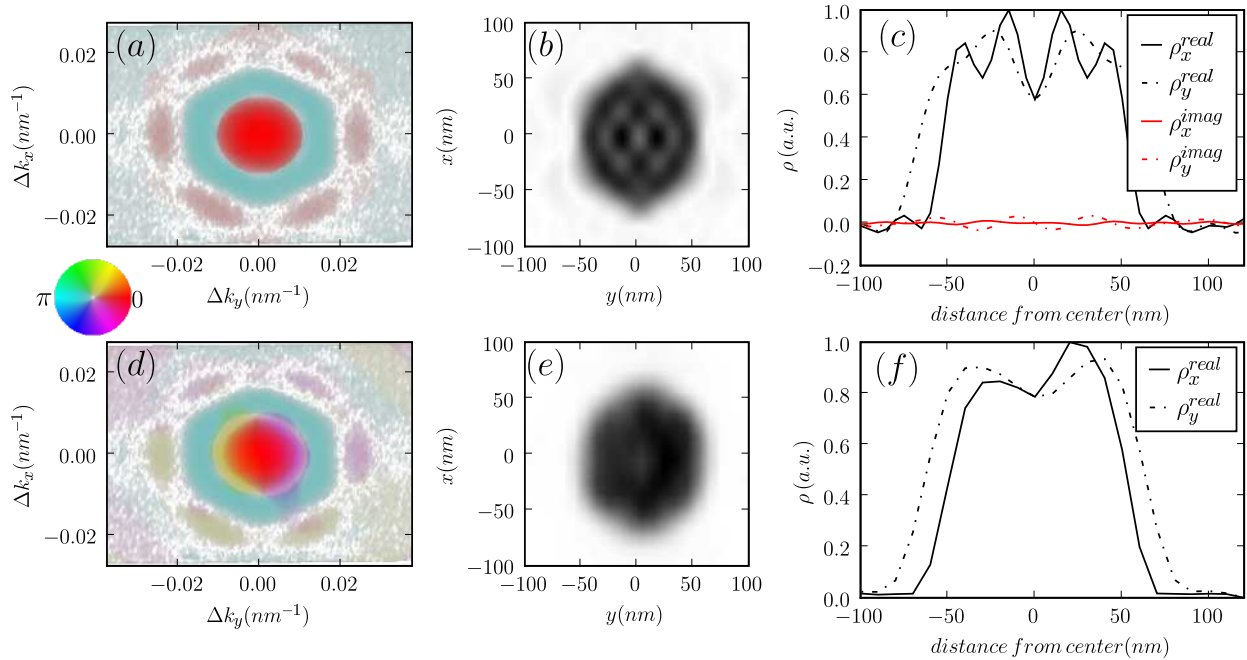}
\caption{\label{fig:fig3}(color online) The real shape (the projection of the electronic density) of the NW can be reconstructed from the measured scattered amplitude by recovering the lost phase information, either by \textit{predicting} the phase values (a-b-c) or by using \textit{iterative phase retrieval algorithms} (d-e-f) (see text for details) (a) projection of the experimental scattered amplitude, with model phases - the saturation corresponds to the logarithm of the intensity, and the color to the phase. Relative coordinates around the (111) reflection are given in $k=2\sin\theta/\lambda$ units. (b) NW density cross-section, obtained by inverse Fourier transform of (a) ; the distance between opposite facets is equal to $\approx 95\,nm$. (c) density profiles along the horizontal ($\rho_x$) and vertical ($\rho_y$) directions, which demonstrate (the real components being positive and the imaginary one negligeable) that the choice of alternating (+) and (-) sign for successive rings is correct. The dips of the density inside the NW are due to the limited extent of the intensity scattered by the silicon NW.(d) Experimental amplitude with the phase recovered using \textit{ab initio} Fourier-recycling algorithms. (e) the best reconstruction obtained and (f) the corresponding cross-sections (positivity is imposed in the algorithm).}
\end{figure*}

  In our case the recorded intensity, shown in Fig.~\ref{fig:fig2}, exhibits a clear sixfold symmetry which suggests that the original $<111>$ wire was hexagonaly-shaped, as exhibited on most wires by scanning electron microscopy (SEM). Furthermore, the calculation of the FT of an hexagon shape reveals that the phase is either $0$ or $\pi$ (due to the presence of a center of symmetry), as shown in Fig.~\ref{fig:figSIMU}, and presents concentric hexagonal-shaped rings with \textit{alternated $0/\pi$ phases}. In order to reconstruct the shape of the NW 2D cross-section\footnote{ As the NW is much longer (several $\mu m$ vs. $\approx 100\,nm$), the FT of its shape is essentially 2D.}, we projected the recorded intensity onto the plane perpendicular to the wire and applied an alternated sign ($+1/-1$) to the concentric rings (see Fig.~\ref{fig:fig3}a) and then computed the inverse FT to yield the shape of the NW, which can be seen in Fig.~\ref{fig:fig3}b.

  The obtained shape of the NW exhibits the expected hexagonal symmetry. In order to check that the 'alternating $0/\pi$ phases' was the correct solution to the lost phases, we checked (see Fig.~\ref{fig:fig3}c) that the real part of the inverse FT was strictly positive inside the object, and much larger than the imaginary part (ratio $\frac{Real}{Imag}>50$), which validates the phase choice. Note that this outcome was not guaranteed, as the original data was not \textit{strictly} centrosymmetric, and no attempt (\textit{e.g.} by symmetry averaging) was made to modify this.

  The size (FWHM distance between opposite faces) of the wire is $\approx\,95\,nm$, with a  resolution of about $28\,nm$ (calculated from $\Delta k_{max}=0.035nm^{-1}$) for the reconstruction ; the NW appears to be slightly elongated in one direction, which may be due to an asymmetric NW growth, probably due to the original seed combined with kinetic limitations. In the extracted density map shown in Fig.~\ref{fig:fig3}b, we observe a variation amounting up to $20\%$ from the average value. This is very likely due  to the limited extent of the recorded intensity: only three interference orders (0,1,2) have been recorded, due to the small size of the NW and the weak scattering power of the silicon atoms, which both limits the real-space resolution and can create ripples inside the reconstructed density ; a simulation of the effect of truncated data is presented in Fig.~\ref{fig:figSIMU} and has allowed us to reproduce the electronic density dips (Fig.~\ref{fig:figSIMU}d). Another source for the limited resolution on the sides of the NW comes from the zig-zag nature of the NW facets (see  Fig.~\ref{fig:fig1}), which blurs the average density on the NW borders.

  In order to check that the reconstructed structure using alternating phases was the best possible, we also performed an \textit{ab initio} reconstruction using a combination of the error-reduction, hybrid input/output and charge flipping algorithms, using finite support and positivity as constraints, following the process depicted in Ref. \cite{wu05}. The best set of phases that we obtained\footnote{The optimization used 200 to 400 cycles, with a fixed square support more than twice the size of the NW, and was repeated 400 times to avoid stagnation.} (see Fig.~\ref{fig:fig3}d,e,f) has an R-factor\cite{wu05} equal to $0.7\%$, features the expected alternating phases, and the reconstructed NW shape also presents a quasi-hexagonal shape, thus confirming the results of the inversion using "predicted phases".

  One expected problem when studying single, nano-sized objects using an X-ray beam is radiation damage: this is a well known issue for macromolecular compounds \cite{marchesini03,murray04}, with a dose limit estimated to $2.10^7\,Gy$ ($1\,Gy=1\,J/kg$). In the case of inorganic compounds there is no acknowledged dose limit, as the compounds do not present the specific weak bonds ($-CO_2$, $S-S$) that are likely to break under irradiation.

\begin{figure}
\includegraphics[width=86mm]{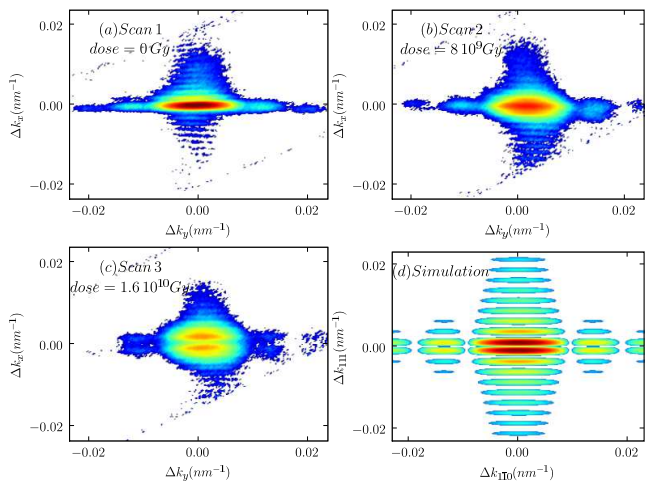}
\caption{\label{fig:fig4}(color online) Effect of radiation damage: (a,b,c) evolution of the 3D diffraction pattern of a single Si NW around the $(111)$ reflection for successive scans, with a total exposure time of $4000\,s$ per scan (a fourth scan is not presented here). The projection is presented in the direction perpendicular to the NW axis, using a logarithmic colour scheme. The short vertical oscillations are due to the finite length of the wire, and almost disappear in later scans, due to the damage to the NW. The fringes which can be seen parallel to the diagonal of the image correspond to the FT of the NW cross-section, as seen in Fig.~\ref{fig:fig2} for another wire. The split of the main diffraction peak could be due to the breaking of the NW: (d) simulated scattered amplitude around the $(111)$ reflection of a broken wire, with two parts with an equal length of 400 nm along $<111>$ and a width of 105 nm along $<1\overline{1}0>$, with a separation between the two parts equal to half the interplanar distance $d_{111}$. This separation leads to two maxima at the $(111)$ reflection position.} 
\end{figure}

  We have however observed that the NW can break under the X-ray beam, as is illustrated in Fig.~\ref{fig:fig4}. We measured the scattering from a single wire by rotating the sample over a $0.8^{\circ}$ angular range with $0.02^{\circ}$ steps, and a $50\,s$ exposure time per image. This was repeated four times in order to accumulate more statistics, and the evolution of the projection of the 3D recorded scattering is shown in Fig.~\ref{fig:fig4}. During this measurement, the flux of the experiment was equal to $\approx 4\,ph/s/\AA^2$. Given the silicon absorption cross-section $\sigma_{E=10keV}=1.5\times10^{-5}\AA^2$, this results in a absorbed power per atom of $0.6\,eV/s$, or equivalently $2\times10^6\,Gy/s$, \textit{i.e.} a total dose of $8\times10^{9} Gy$ per scan.

  During the experiment, the absorbed energy is not accumulated and can be evacuated through fluorescence, thermal radiation or conductivity, but point defects will nevertheless be created in the sample. The presence of already existing defects (such as a stacking fault -a common occurence in NW \cite{verh06}, point defects or dislocations) can however be considered as a "weak" part of the structure, which could lead to a broken wire as exhibited in Fig.~\ref{fig:fig4}. However new experiments (e.g. with a more focused and intense beam) would be necessary to quantify the relationship between the dose, existing faults, and the effects of radiation damage to a NW.

  In conclusion, we have demonstrated that it was possible to use coherent diffraction imaging on single $\approx100\,nm$ NW to recover their shape with a $\approx\,28\,nm$ resolution, even in the case of weak scatterers like silicon. Evidently the interest of this result does not lie in the reconstruction of the shape of homogeneous NW, since scanning and transmission electron microscopy routinely yield the external shape of nano-sized object with a better resolution. However this technique will be particularly useful in the case of \textbf{heterogeneous NW}, \textit{i.e.} with longitudinal or radial (core-shell) heterostructures: these type of NW exhibit strain fields and chemical gradients at the interface for which X-ray diffraction remains the most sensitive technique, and will be the focus of upcoming experiments. Such an experiment would allow to recover the 3D complex field describing both the density and the distortion of the lattice, as depicted in Ref \cite{pfeifer06}.  

 Moreover constant progress is made for X-ray optics in order to improve the coherence and flux on synchrotron beamlines - e.g. as of the writing of this article the focused beam size on the ESRF ID01  beamline is much smaller ($10^{10}\,ph/s$ in $2.5\times0.5\,\mu m^2$ ), yielding a flux 20 times more intense than during our experiment, which should allow to study smaller NWs with a better resolution.

This work has been partially performed under the EU program NODE 015783. The authors would like to thank the ID01 ESRF beamline for their technical help, and acknowledge Prof. Ian Robinson for helpful discussions.




\bibliography{2007CDI-NanoWire}

\end{document}